SHARAD Illuminates Deeper Martian Subsurface Structures
with a Boost from Very Large Rolls of the MRO Spacecraft

Nathaniel E. Putzig[1], Gareth A. Morgan[2], Matthew R. Perry[1],
Bruce A. Campbell[3], Jennifer L. Whitten[3], Fabrizio Bernardini[4],
Alessandro DiCarlofelice[5], Piero Tognolatti[5], Pierfrancesco Lombardo[6]

ABSTRACT

Throughout its mission, the Mars Reconnaissance Orbiter (MRO) has often rolled about its along-track axis by up to 28° to partially compensate for the suboptimal location of the Shallow Radar (SHARAD) antenna along an edge of the spacecraft that is opposite the imaging payload deck, thereby enhancing the signal-to-noise ratio (S/N) of echoes returned from the surface. After recent modeling work predicted that a much larger roll would improve the S/N by ~10 dB relative to nadir-pointed observing, MRO began a limited series of 120° roll maneuvers to test the effects on radar sounding. Three such SHARAD very-large-roll (VLR) observations have been acquired since May 2023, and they show dramatic improvements in signal clarity and depth of penetration, with S/N increasing by 9, 11, and 14 dB over that of nearly coincident observations at 0° roll angle. In low dielectric terrains, the first and second VLR observations enabled basal detections at depths previously unachievable, reaching depths of 800 m in Medusae Fossae materials and 1500 m through the ice of Ultimi Scopuli, respectively. The second VLR observation also obtained enhanced reflections throughout the ice stack. In the higher dielectric terrain of Amazonis Planitia, the third VLR observation improved continuity of a dipping subsurface interface, but it revealed neither an extension of the interface to greater depths nor any deeper interfaces. The MRO mission intends to obtain more SHARAD VLR observations of polar terrains and of midlatitude glacial and ground ices, sediments, and volcanics.

---

[1] Planetary Science Institute, Lakewood, CO, USA; Corresponding author nathaniel@putzig.com
[2] Planetary Science Institute, Tucson, AZ, 85719, USA
[3] Smithsonian Institution Center for Earth & Planetary Studies, Washington, DC, USA
[4] INAF/IAPS Istituto di Astrofisica e Planetologia Spaziale, Roma, 00133, ITALY
[5] Università degli Studi dell'Aquila, L'Aquila, 67100, ITALY
[6] Sapienza Università di Roma, Roma, 00133, ITALY

1. INTRODUCTION

Along with the other instruments onboard the Mars Reconnaissance Orbiter (MRO), the Shallow Radar (SHARAD) instrument (Seu *et al.* 2007) began acquiring science data at Mars in the autumn of 2006. SHARAD is a 10 W sounding radar that emits frequency-modulated pulses, downswept from 25 to 15 MHz and providing a 15 m free-space range resolution with a Fresnel zone at the Martian surface that yields 3–6 km lateral resolution. The along-track resolution is improved 10-fold through synthetic aperture processing. Due to downlink bandwidth limitations, SHARAD is a targeted instrument whereby the instrument team chooses discrete observations over geographic regions of interest over which the spacecraft passes and ranging in extent from 30 km to 6000 km. As of December 2024, over 37,000 observations have been acquired covering 60% of Mars at a nominal 3-km track-width resolution. SHARAD data have contributed to a broad range of discoveries over the course of the MRO mission, revealing the nature of materials and geologic structures in the polar caps and in mid-latitude terrains containing debris-covered glaciers, near-surface ground ice, and stacks of volcanic and sedimentary materials. See Putzig *et al.* (2024) for a summary of findings.

Due to the limited availability of space on MRO's imaging payload deck and concerns about the potential for electromagnetic interference, it was necessary to place SHARAD's antenna along an edge of the opposite side of the spacecraft (**Figure 1**). While it was understood that this deployment would impact the power of SHARAD's transmitted and received signals, only limited calibration of the effects of spacecraft and appendage orientation was possible prior to the August 2005 launch and upon arrival in orbit at Mars (Croci *et al.* 2011). Driven primarily by the need for the imaging cameras to observe targets off nadir, MRO standard operations were established with an allowance for observing at roll angles up to ±30° about the along-track axis, which is parallel to SHARAD's 10-m dipole antenna. To assess the effects of spacecraft and appendage orientation on SHARAD signals, the calibration exercises included acquisition of data with the communications high-gain antenna (HGA) and solar arrays (SA) in various orientations and over a range of roll angles. It was found that, relative to observing with the HGA pointed in the positive cross-track direction,

*Putzig et al. SHARAD VLR Results - For submission to Planetary Science Journal*the SA oriented parallel to the along-track axis and facing away from Mars, and the spacecraft oriented to nadir (0° roll), the SHARAD round-trip gain ranged from -3.4 dB to +3.8 dB for various other appendage orientations (primarily driven by the SA) and from -1 dB to +3 dB for roll angles transitioning from -25° to +25° (Croci *et al.* 2007, 2011). A more complete analysis of the dependence of system gain on SA and HGA positions, and for limited rolls, was achieved after more than a decade of observations (Campbell *et al.* 2021a). Empirical correlations suggest about a 4-dB total range of SHARAD round-trip gain with the spacecraft in the nadir-pointing geometry, with an additional gain of up to 6 dB possible for roll angles close to 30°. As a result of the earlier tests, the mission established a standard SA orientation to use for SHARAD observing on the nightside

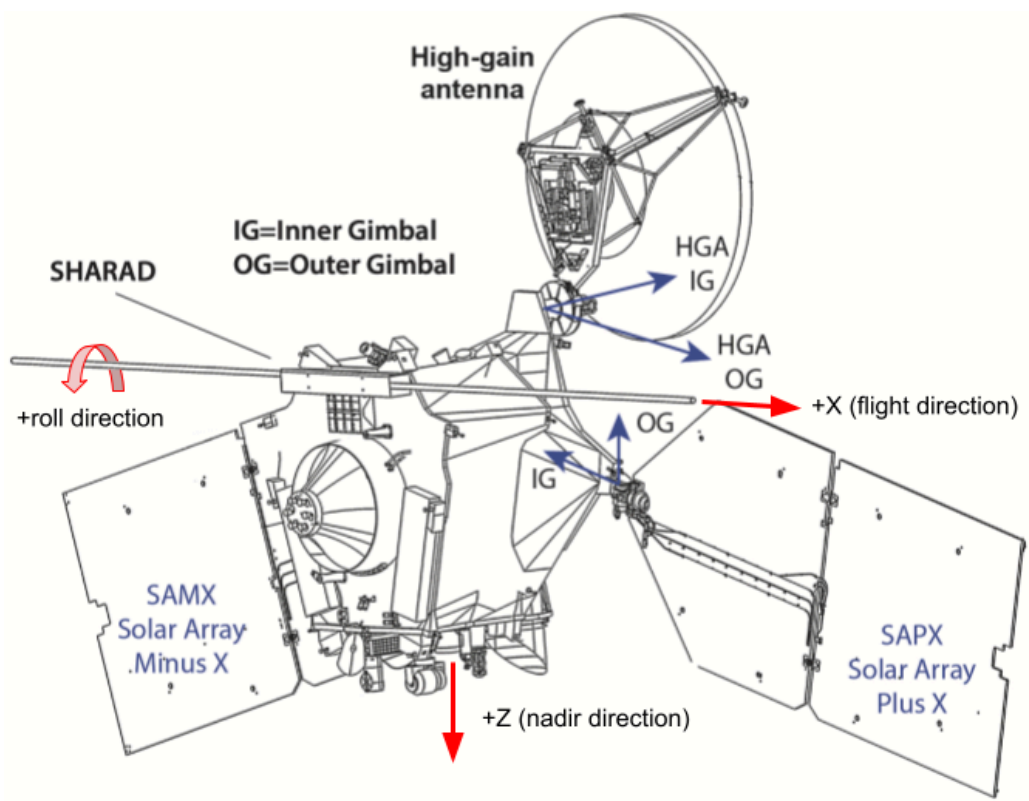

**Figure 1.** *MRO's high gain antenna (HGA) and solar array (SA) orientations and gimbal rotation axes. Spacecraft rolls occur about the X axis, which represents the flight direction and is parallel to the SHARAD dipole antenna. The Z axis and nominal nadir (planet-facing) direction is toward the bottom of the figure. Image credit: NASA. Modified after Figure 7 of Campbell et al. (2021a).*

ACCEPTED MANUSCRIPT          3 of 22          DOI:10.3847/PSJ/addbe1

wherein the SA are folded back and away from the SHARAD antenna, the so-called "SHARAD park" position (higher SA inner gimbal angles provide greater overall gain at lower roll angles). This position is only used when the spacecraft is in eclipse, as the SA otherwise tracks the Sun for energy collection. At the time, it was realized that additional gain would occur at roll angles higher than 25°, and the majority of SHARAD rolled observations over the course of the mission have been carried out at roll angles approaching the 30° limit. It was not foreseen that operation of SHARAD at roll angles above that limit would be feasible, and no effort to assess the effects above a +30° roll angle was pursued until recently.

In subsequent sections, we summarize recent work by DiCarlofelice *et al.* (2024) to extend the modeling of the effects of spacecraft and appendage orientation, we describe operational considerations and planning efforts to acquire SHARAD observations at very large roll (VLR) angles, and we present results from the first three such observations. For each result, we include an assessment of the improvement in signal-to-noise ratio (S/N), with details provided in the Appendix. We then discuss plans for continued use of this new VLR observing mode over various targets of interest during MRO's seventh extended mission from October 2025 through September 2028.

## 2. RECENT ANTENNA MODELING

The MRO spacecraft and its instruments are in remarkably good condition after more than 18 years in operation at Mars (Zurek *et al.* 2024). With each two- or three-year extended mission, the MRO Project aims to broaden its capabilities to pursue a renewed set of science objectives. In its sixth extended mission (October 2022 through September 2025), MRO's goals have come to include improving SHARAD's signal strength and depth of penetration by considering spacecraft rolls beyond the nominal 30° limit.

Given the lack of characterization of the SHARAD antenna at high roll angles, the SHARAD team began a collaboration with researchers at the University of L'Aquila (authors Tognolatti and DiCarlofelice) with expertise in electromagnetic modeling of radar systems. A new method-of-moments numerical model was developed, one with the ability to predict the SHARAD dipole antenna pattern for any combination of MRO attitude, SA orientation, and HGA orientation (DiCarlofelice *et al.* 2024). While most spacecraft elements were modeled as perfect electrical conductors, the SA were treated as multilayered structures with varying properties to better account for their effects on SHARAD signals. Comparisons between the model results and the empirical analysis of the influence of the HGA and SA orientation on the strength of the SHARAD surface return at nadir (Campbell *et al.* 2021a) demonstrated good agreement, adding weight to the modeling approach. See DiCarlofelice *et al.* (2024) for details of the material and electrical property assumptions used in the model.

Results from the antenna modeling confirm that the primary factors affecting SHARAD S/N are the spacecraft roll angle and the orientation of the SA panels, with little to no effect from the orientation of the HGA. In addition, the effects on signal gain from rotations of the SA about their longitudinal axes (i.e., using outer gimbals) were found to be relatively minor, so analysis efforts were concentrated on the effects of SA rotations that occur largely about the spacecraft Z axis (i.e. using inner gimbals), which is oriented parallel to the nominal nadir direction (**Figure 1**). Varying somewhat with the radar frequency, a minimum in the radiation pattern is found at a roll angle of -15° when the SA gimbal angles are all set to 0°, where the SA panels are parallel to and facing away from the SHARAD antenna (as in **Figure 1**), and the gain remains low at a roll angle of 0°. In practice, the effect at 0° is mitigated somewhat by setting the gimbal angles to the SHARAD park position for nightside observing at roll angles below 30° (**Figure 2a**). In contrast, the modeling shows a maximum gain at a roll angle of 120° when the SA gimbal angles are all set to 0°, so the SHARAD park position is not used for VLR observations (**Figure 2b**).

## 3. OPERATIONAL CONSIDERATIONS

While the MRO Project limits the spacecraft roll angle to 30° or less during normal spacecraft operations, there have been a number of instances where more-extreme off-nadir pointings have been undertaken during the course of the mission. Until 2023, these special cases have all been for the use of the High Resolution Imaging Science Experiment (HiRISE) to observe targets that include the Earth–Moon system, the Martian moons

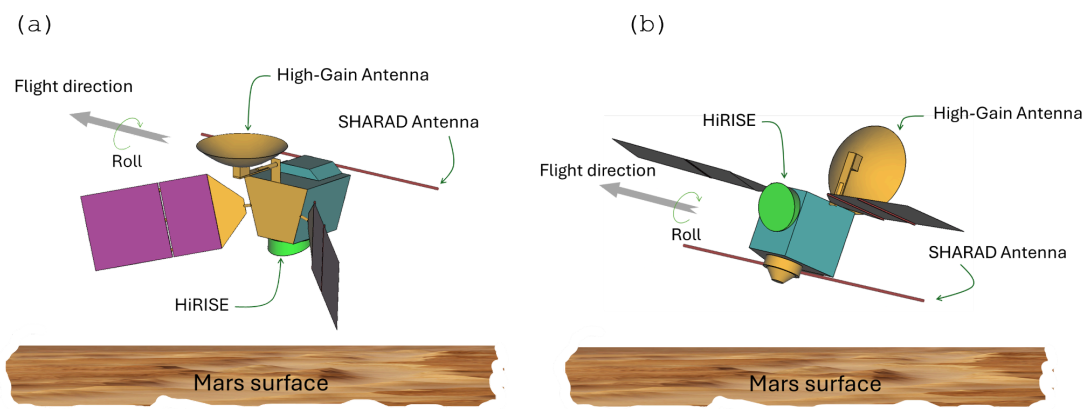

***Figure 2.*** *Nominal orientations for Mars Reconnaissance Orbiter when SHARAD is observing at (a) 0° roll angle with SA in SHARAD park position and (b) 120° roll angle with SA gimbals in all-zeros position. Modified after Figure 1 of DiCarlofelice et al. (2024).*

Phobos and Deimos, Jupiter, comets ISON and Siding Spring, and landing spacecraft on descent to the surface (Phoenix, Curiosity, and Perseverance). Rolls outside of the normal limit require careful handling by the MRO planning teams, as they could impact the ability of the spacecraft to recharge its batteries or risk inadvertently pointing cameras or other thermally sensitive devices toward the Sun, potentially damaging them. For the 120° roll desired by SHARAD, the spacecraft is essentially turned completely over, and these concerns are elevated. To better avoid any impacts on energy management and other instruments, the SHARAD VLR observations have all been taken while the spacecraft is in eclipse on the nightside of Mars. During normal operations, rolled observations are executed with the spacecraft gradually rolling into the desired roll angle at the center of each observation and then gradually rolling back to a nadir orientation without pausing during the observation. For the VLR case, the spacecraft rolls to the desired 120° roll angle before the start of the observation, then holds at that orientation throughout the observation before rolling back to the nadir orientation.

As discussed above, the positions of the SA, due to their influence on the SHARAD signal gain, also need to be considered. Prior to a VLR observation, the SA are repositioned to the SHARAD favorable all-zero gimbal angle orientation, and they are held in that orientation for the duration of the observation. Upon completion of the observation, the gimbal angles are reoriented for dayside energy collection. As an engineering precaution, the spacecraft roll and SA movement have been sequenced to occur in series rather than simultaneously, resulting in time blocks of up to 16 min for reorientation and return to nominal condition. In addition, every SHARAD observation requires a 3 min set-up time, which has to occur after the spacecraft and SA movements. Depending on the circumstances for a particular observation, some or all of the spacecraft rolling, the SA reorientation, and the set-up period may need to occur also within the spacecraft eclipse period, placing significant seasonal constraints on the latitudes available for VLR observing.

Because of the complex nature of the VLR observations, MRO has only been able to acquire three of them over the two years since planning the first one. The MRO Project is working toward a standardization of the procedure that would allow more frequent use of this new observing mode (on the order of 10 or 12 times per year). Current discussions involve confining all of the maneuvers for such standard VLR observations to the spacecraft eclipse period, which would limit them to a relatively narrow band of latitudes that evolves with the seasons (**Figure 3**). This scheme would allow access to all mid-latitude terrains over the course of a Martian year, but the polar regions would not be accessible. Non-standard, specialized planning (involving some maneuvering outside of eclipse) as was done for the second VLR observation would continue to be necessary to access the polar regions.

## 4. SHARAD OBSERVATION RESULTS

The first SHARAD VLR observation (7858301 or VLR1 hereinafter) was acquired on May 2, 2023, extending from 13.28°N, 199.10°E at the southern edge of Amazonis Planitia, across Eumenides Dorsum of the Medusae Fossae Formation, and into Terra Sirenum in the southern highlands at 36.04°S, 192.84°E (**Figure 4**). As predicted, a substantial boost in S/N was achieved relative to that obtained at lower roll angles, which is apparent upon a visual inspection of the radargrams[7] (**Figures 4c–4e**) and can be assessed and compared quantitatively along-track (**Figure 4b**). Here, the mean S/N improvement is 9 dB between 0° and 120° roll observing. Where they cross the western fringes of Eumenides Dorsum, the SHARAD observations exhibit subsurface reflections from the base of the Medusae Fossae Formation, and these reflections exhibit greater power and extend to greater depths in the VLR1 observation. More broadly, that observation also exhibits clutter signals from much more distal sources than its counterparts acquired at lower roll angles, and that clutter even extends beyond the 45-km cross-track cut-off distance normally used in the generation of cluttergrams[8] (Christoffersen *et al.* 2021)(compare **Figures 4c** and **4f**).

---

[7] Radargrams are profile images showing the returned radar power with delay time on the vertical axis and along-track distance on the horizontal axis.

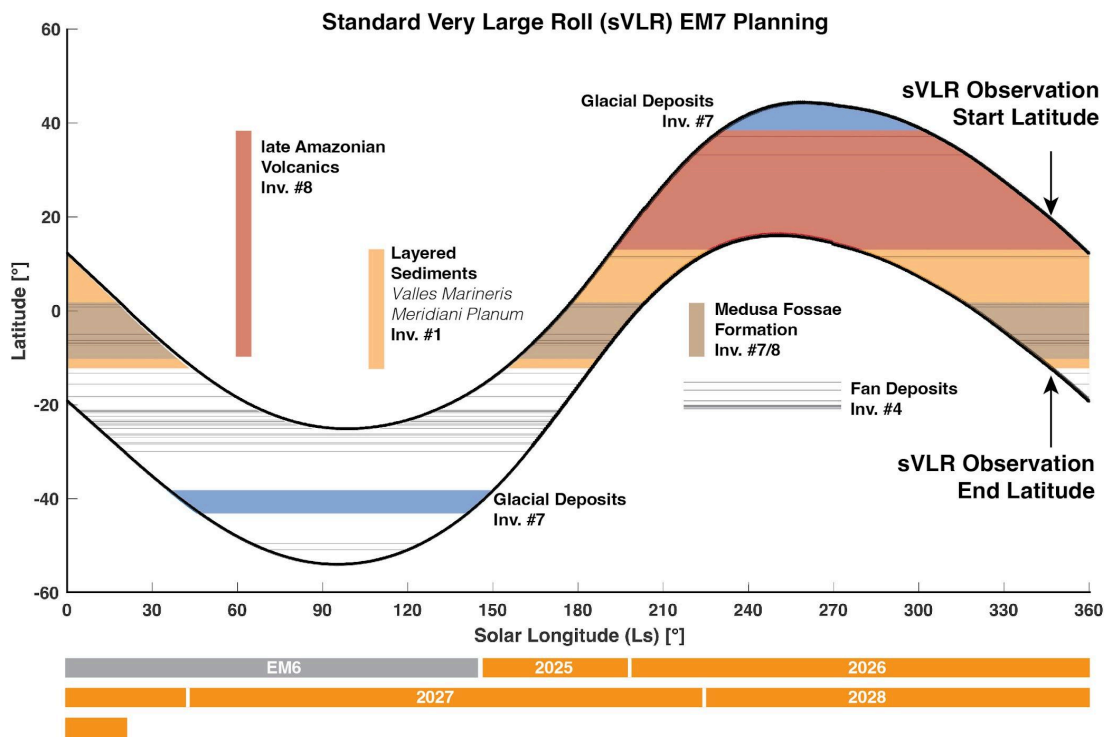

***Figure 3.*** *Planning diagram showing variation in latitude range (between heavy black lines) vs. season (solar longitude, $L_S$) for SHARAD standard VLR observations wherein all associated spacecraft maneuvers occur when the MRO spacecraft is in eclipse. Colors and lighter black lines show accessible terrains, keyed to investigations proposed for MRO Extended Mission 7. Polar regions are not accessible under these constraints.*

---

[8] Cluttergrams are simulated radargrams that show signals expected to be generated solely from surface reflections, where off-nadir surface reflections are termed "clutter".

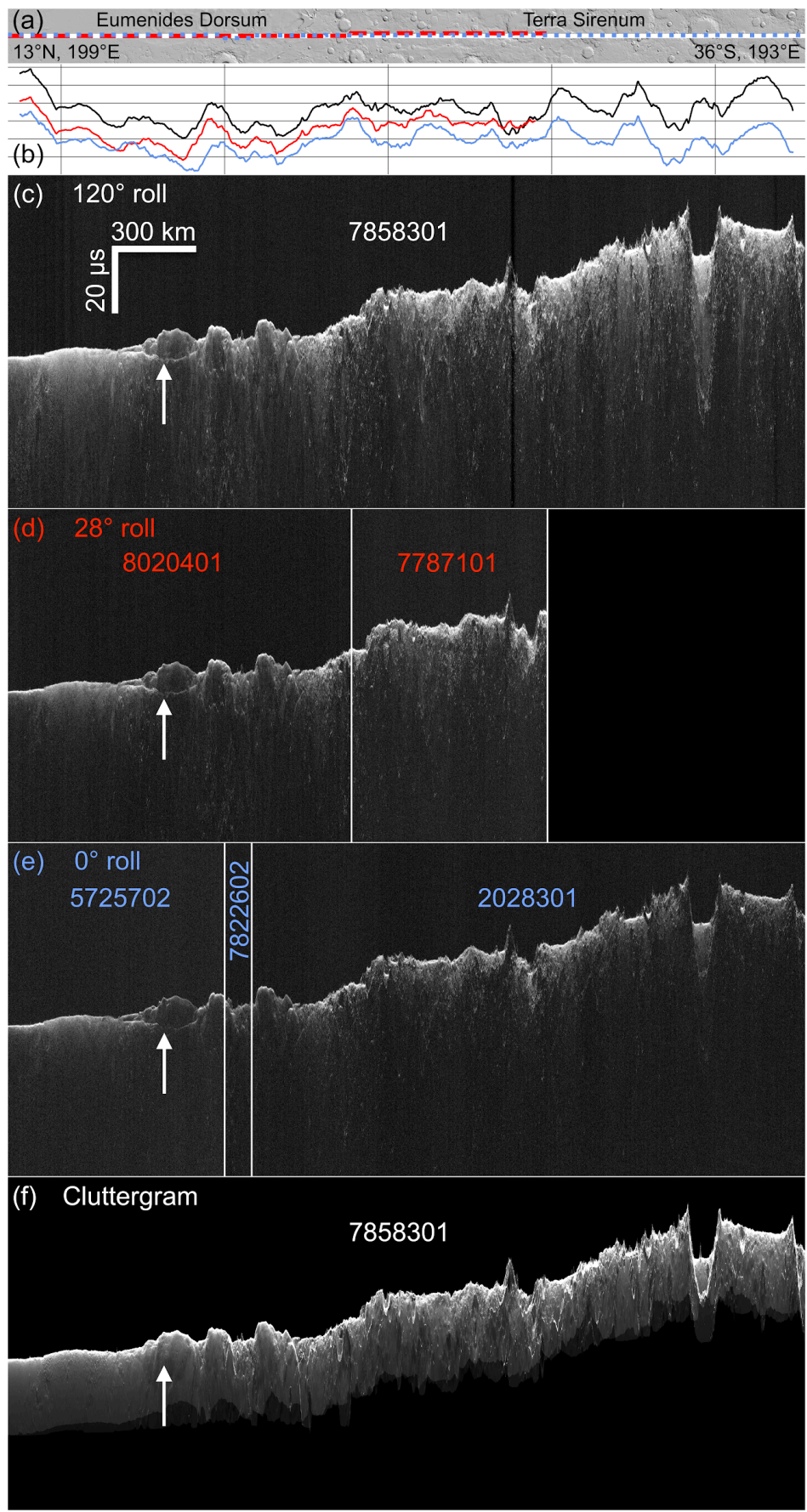

*Figure 4. Comparison of the first very-large-roll SHARAD observation (VLR1) with nearly coincident portions of other observations taken at lower roll angles. (a) Strip map showing the topography along the ground tracks of the SHARAD observations (MOLA shaded relief), with the white center line corresponding to VLR1, the red dashed lines corresponding to observations 8020401 and 7787101, and the blue dotted lines corresponding to observations 5725702, 7822602, and 2028301. (b) Estimated along-track S/N for VLR1 (black line), observations 8020401 and 7787101 (red line), and observations 5725702, 7822602, and 2028301 (blue line). S/N values range from 8 dB to 39 dB. The mean difference between the black and blue lines is 9 dB. (c) Radargram for VLR1 acquired at 120° roll angle. (d) Radargrams for segments of observations 8020401 and 7787101 acquired at ~28° roll angle. (e) Radargrams for segments of observations 5725702, 7822602, and 2028301 acquired at 0° roll angle. (f) Cluttergram for VLR1. Power in the radargrams is scaled from -3 dB to 24 dB. White arrows in (c–f) show where VLR1 achieved greater depth of penetration than in other observations. The lack of a corresponding feature in the cluttergram provides confidence in a subsurface source of the reflection.*

The second SHARAD VLR observation (7983501 or VLR2 hereinafter) was acquired on August 8, 2023, extending from 69.98°S, 204.88°E at the southern edge of Terra Sirenum, across a partially filled set of overlapping craters on the edge of Planum Australe, and into the Ultimi Scopuli region of Planum Australe at 84.67°S, 181.19°E (**Figure 5**). The SHARAD team planned this observation to cross the area in Ultimi Scopuli where the Mars Express Mars Advanced Radar for Subsurface and Ionosphere Sounding (MARSIS) has obtained very strong basal reflections that have been interpreted as evidence for subglacial liquid water (Orosei *et al.* 2018; Lauro *et al.* 2021). Prior efforts to observe the same feature with SHARAD using roll angles approaching the normal limit did not achieve sufficient depth of penetration (e.g. **Figures 5d** and **5e**). Due to planning constraints, the VLR2 ground track does not cross the area of the primary MARSIS feature mapped by Orosei *et al.* (2018) although it does pass through the region containing outlying features with high basal power as mapped by Lauro *et al.* (2021). In any case, the boost in S/N is once again very evident upon inspection of the radargrams, with a mean improvement of 11 dB between 0° and 120° roll observing. In this case, the improved depth of signal penetration extends SHARAD's capabilities by ~500 m to the overall depth of the base of ice at ~1500 m.

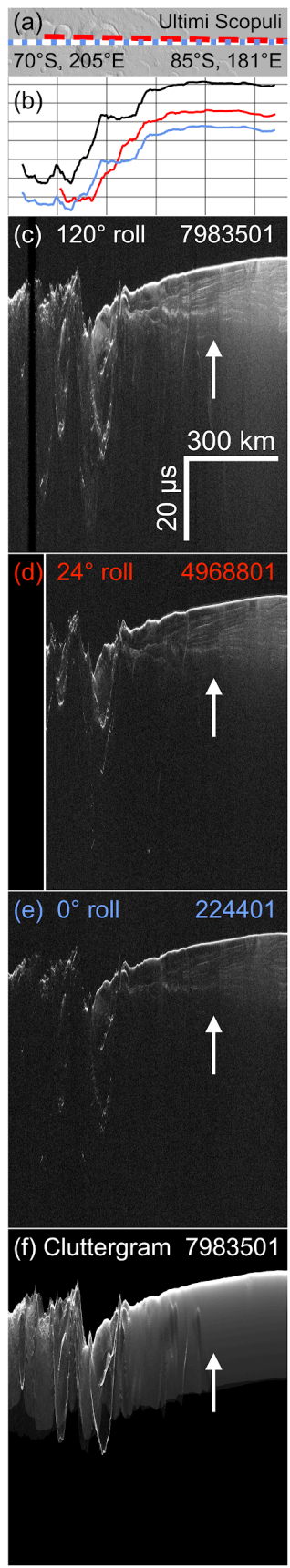

**Figure 5.** *Comparison of the second very-large-roll SHARAD observation (VLR2) with nearly coincident portions of other observations taken at lower roll angles. (a) Strip map showing the topography along the ground tracks of the SHARAD observations (MOLA shaded relief), with the white center line corresponding to VLR2, the red dashed line corresponding to observation 4968801, and the blue dotted line corresponding to observation 224401. (b) Estimated along-track signal-to-noise ratio for each SHARAD observation segment. S/N values range from 5 dB to 42 dB. The mean difference between the black and blue lines is 11 dB. (c) Radargram for VLR2 acquired at 120° roll angle. (d) Radargram for segment of observation 4968801 acquired at ~24° roll angle. (e) Radargram for segment of observation 224401 acquired at 0° roll angle. (f) Cluttergram for VLR2. Power in the radargrams is scaled from −3 dB to 24 dB. White arrows in (c–f) show where VLR2 achieved greater depth of penetration than in other observations. The faint reflection just above the white arrow in Figure 5c occurs at the same delay time where MARSIS obtained strong basal reflections. The lack of a corresponding feature in the cluttergram provides confidence in a subsurface source of the reflection.*

The third SHARAD VLR observation (8503001 or VLR3 hereinafter) was acquired on September 15, 2024, extending from 41.83°N, 199.54°E at the southern edge of Arcadia Planitia, across central Amazonis Planitia to the west of Eumenides Dorsum, and into the northern edge of Terra Sirenum at 8.12°S, 193.01°E (**Figure 6**). The SHARAD team planned this observation to test whether the improved S/N would extend to greater depths a dipping reflector seen in prior observations of Amazonis Planitia that has been interpreted to be the interface between sedimentary deposits of the Vastitas Borealis Formation and underlying Hesperian volcanic plains (Campbell *et al.* 2008). The team also wanted to see whether additional interfaces would be found either at greater depths below the one already identified or in other areas along the ground track. VLR3 achieved the greatest boost in S/N of the three VLR observations, with a 14 dB increase between 0° and 120° roll observing. In similar fashion to the others, VLR3 obtained much more distal clutter signals, but it produced no geographic extension of existing reflectors nor any clear evidence of greater depth of penetration.

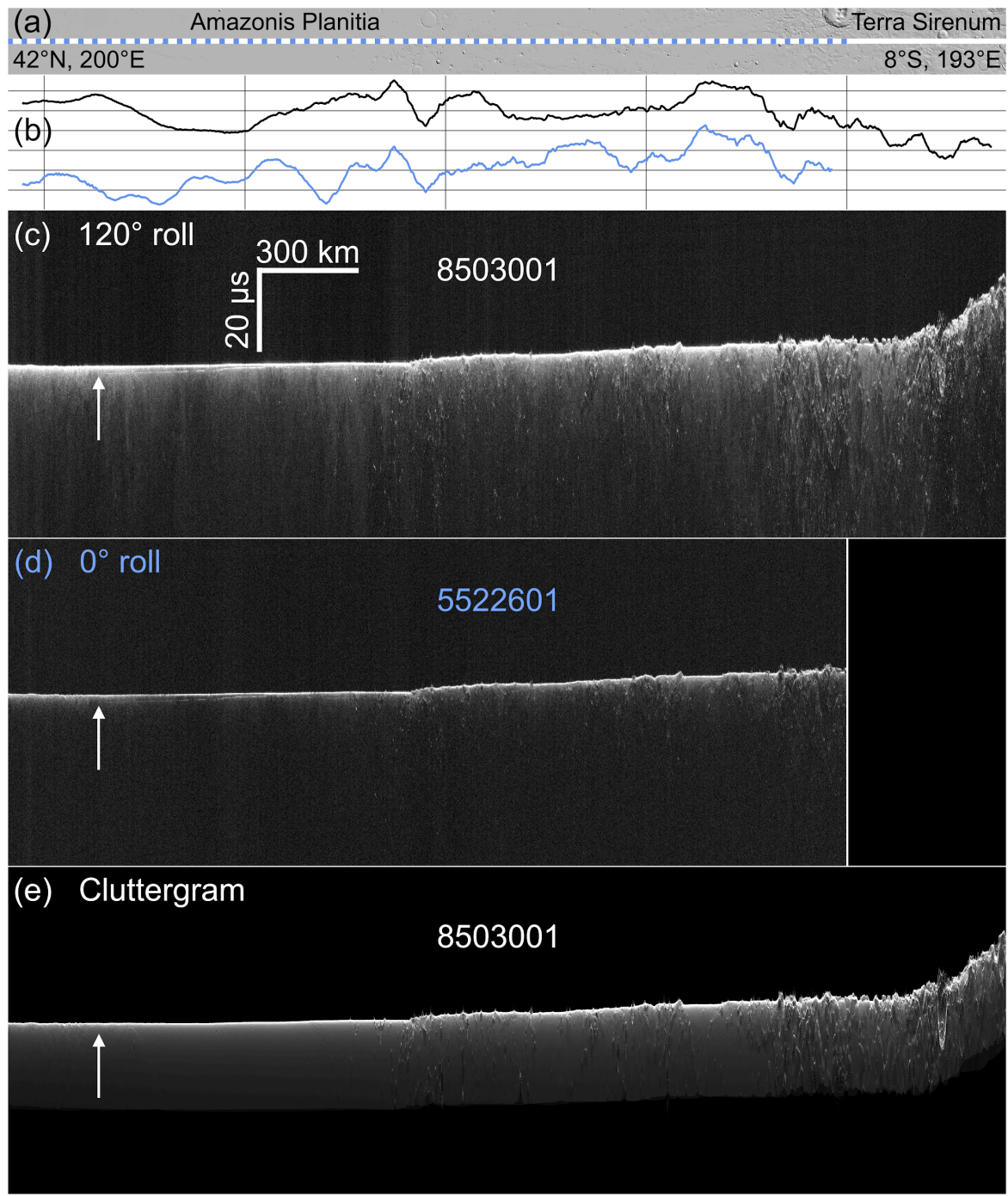

***Figure 6.*** *Comparison of the third very-large-roll SHARAD observation (VLR3) with a nearly coincident portion of another observation taken at 0° roll angle. (a) Strip map showing the topography along the ground tracks of the SHARAD observations (MOLA shaded relief), with the white center line corresponding to VLR3 and the blue dashed line corresponding to observation 5522601. (b) Estimated along-track signal-to-noise ratio for each SHARAD observation segment. S/N values range from 7 dB to 41 dB. The mean difference between the black and blue lines is 14 dB. (c) Radargram for VLR3 acquired at 120° roll angle. (d) Radargrams for segment of observation 5522601 acquired at 0° roll angle. (e) Cluttergram for VLR3. Power in the radargrams is scaled from -3 dB to 24 dB. White arrows in (c–e) show the northernmost extent of a subsurface reflection that appears in both observations. The lack of a corresponding feature in the cluttergram provides confidence in a subsurface source of the reflection.*

## 5. DISCUSSION

The SHARAD VLR observations demonstrate that the prior antenna modeling effort accurately predicted the improved S/N achievable with the 120° roll and optimal SA orientation of the MRO spacecraft. The three VLR observations acquired to date all show improved clarity for subsurface reflections, and for VLR1 and VLR2 the signal improvement resulted in notable increases in the depth of penetration. These deeper interfaces had been previously illuminated by the MARSIS instrument, albeit at much coarser resolution. The same is not true for the dipping interface in Amazonis Planitia that is seen in the SHARAD observations, and this interface is significantly shallower than the others. At depths of 100–170 m (Campbell *et al.* 2008), that interface is within one or two resolution cells of the surface for MARSIS frequencies, which likely explains the lack of detection by that instrument.

Landforms of the Medusae Fossae Formation have long been of interest to radar sounding studies, as their low bulk dielectric subsurface properties place constraints on their emplacement and the extent to which they may harbor deep subsurface ice (Watters *et al.* 2007; Carter *et al.* 2009; Morgan *et al.* 2015; Campbell *et al.* 2021b). While MARSIS frequencies often allow signal penetration to the base of these materials, the higher-frequency SHARAD signals are typically attenuated within the first several hundred meters. The fact that the VLR1 observation was able to improve the signal penetration depth provides the SHARAD team an incentive to acquire additional VLR observations of these materials, and their equatorial location means that they will be accessible through the standard VLR acquisition process during two Martian seasons ($L_S$ = 315 to 45 and $L_S$ = 160 to 200) (**Figure 3**). One desired outcome is the detection of an internal interface that would support the interpretation of ice being present at several hundred meters depth (Campbell *et al.* 2021b).

The occurrence in Planum Australe of MARSIS basal reflections whose power exceeds that of the surface reflections is a fascinating result, and its explanation as being a result of subglacial liquid water is controversial, sparking a vigorous debate about alternative explanations (Orosei *et al.* 2018; Sori and Bramson 2019; Bierson *et al.* 2021; Khuller and Plaut 2021; Lauro *et al.* 2021; Smith *et al.* 2021; Schroeder and Steinbrügge 2021; Lalich *et al.* 2022; Lauro *et al.* 2023; Lalich *et al.* 2024). As noted by several authors, one difficulty has been the fact that SHARAD observations have previously experienced an attenuation of signal about 500 m above the zone where MARSIS obtains the anomalously powerful reflections, even when acquired with rolls approaching the normal acquisition limit of 30°. The VLR2 observation did obtain reflections from this depth, but their power is quite low. The timing of the opportunity for obtaining the observation did not allow a ground track directly crossing the area of peak MARSIS power, so the low power of the VLR2 reflection cannot speak directly to the liquid water hypothesis. Given the high latitude of the location and the constraints on MRO observing, only a narrow window of time near the Martian southern winter solstice is feasible for such non-standard polar VLR observations, so the mission has had to wait two Earth years before planning to make another attempt in late May of 2025 ($L_s$ = 90 occurs on May 30, 2025).

The fact that no appreciable lateral extension of the Amazonis dipping reflector was achieved with the VLR3 observation came as a surprise, since the main hypothesis to explain its termination in prior observations is the attenuation of SHARAD signals by a change in the overlying material properties (which exhibit high dielectric properties relative to the Medusae Fossae deposit or the ice-rich Planum Australe) and/or elevated scattering due to a change in the surface roughness (Campbell *et al.* 2008). While that explanation remains viable, the degree of the change in properties would need to be severe enough to overcome the 14 dB improvement in S/N afforded by the VLR. An alternative hypothesis is that the termination of the reflector represents a true cessation of the interface at depth (e.g. the edge of an underlying volcanic layer). Similarly, the lack of additional reflections at greater depths may be due to a lack of significant additional dielectric interfaces within a few hundred meters of the surface rather than a loss of signal power.

In considering future use of the non-standard VLR observing mode, a top priority for the MRO Project is to obtain a VLR observation directly over the area of peak MARSIS returns from the base of Planum Boreum in Ultimi Scopuli. The VLR2 result provides a high level of confidence that SHARAD would obtain a reflection from the same feature, and an analysis of the power of that reflection would provide new information to test the liquid water hypothesis. Other areas of Planum Australe that would benefit from VLR observing include regions containing low reflectivity zones and regions impacted by so-called "radar fog" wherein the signals appear scattered as a result of unknown causes (Phillips *et al.* 2011; Whitten and Campbell 2018). In Planum Boreum, VLR observing has the potential to better illuminate the deep interior, especially through the basal units that generally have precluded SHARAD from detecting a base that is typically imaged by MARSIS (Nerozzi *et al.* 2022). A corresponding detection by SHARAD would help to better constrain the material properties of the basal units.

The ongoing development of a standard planning procedure is intended to allow more frequent (~monthly) use of the SHARAD VLR observing mode in the mid-latitudes. Here, the prospect of potentially detecting deeply buried ice in Medusae Fossae materials is especially enticing, as noted above. VLR observing also has the potential to better resolve the bases and any interior layering within the debris-covered glaciers that are prevalent across Deuteronilus and Protonilus Mensae (Petersen *et al.* 2018; Petersen and Holt 2022) and on the eastern Hellas rim (Holt *et al.* 2008). Volcanic terrains also provide good potential targets, notably in Elysium Planitia and Tharsis where SHARAD studies have identified subsurface complexity (Voigt *et al.* 2023; Nerozzi *et al.* 2023; **Simon *et al.* 2014**). **Table 1 presents a list of targets currently under consideration for VLR observing by the SHARAD team.**

Table 1. SHARAD Very Large Roll Observing Targets.

| Target Name | Location | Science Objective |
|---|---|---|
| Basal Units | 80°N, 270°E | Detect cap base and internal layering* |
| Deuteronilus Mensae | 44°N, 25°E | Detect internal layering |
| Protonilus Mensae | 44°N, 45°E | Detect base of glacial ices |
| Tharsis Volcanics | 15°N, 250°E | Probe deeper volcanic interfaces |
| Elysium Planitia | 5°N, 166°E | Probe deeper volcanic interfaces |
| Medusae Fossae | 0°N, 215°E | Detect base and internal layering* |
| Meridiani Planum | 0°N, 358°E | Detect subsurface interfaces* |
| East Hellas Rim | 44°S, 105°E | Detect internal layering |
| Ultimi Scopuli | 81°S, 167°E | Detect cap base and assess power* |
| Ultima Lingula | 83°S, 194°E | Assess radar fog characteristics |

* Target intended to follow up on MARSIS detections.

ACKNOWLEDGMENTS

The authors are grateful to the U.S. National Air and Space Administration (NASA) and Agenzia Spaziale Italiana (ASI) for continuing support of the Mars Reconnaissance Orbiter (MRO) and the Shallow Radar (SHARAD) sounder. We are especially appreciative of the efforts provided by the MRO Project,

Lockheed Martin, and the SHARAD team in enabling the acquisition of VLR observations. ASI provided SHARAD to MRO and leads its operations through a contract to SHARAD Team Leader P. Lombardo at the Sapienza Università di Roma, who was appointed recently to replace R. Seu. U.S. efforts are funded by NASA and the MRO Project via subcontracts from Jet Propulsion Laboratory to Co-Investigator institutions including the Planetary Science Institute and the Smithsonian Institution Center for Earth & Planetary Studies. **The manuscript benefited from constructive comments supplied by two anonymous reviewers.** We dedicate this article to the memory of Roberto Seu, SHARAD's first Team Leader, who passed away in October 2024. His vision for exploration, kindness toward others, and love of family and friends serve as an inspiration to us all.

## DATA AVAILABILITY

SHARAD data are available on the Geosciences node of the Planetary Data System (PDS) at https://pds-geosciences.wustl.edu/missions/mro/sharad.htm. Additional information and products can be found on the SHARAD team website at https://sharad.psi.edu.

## APPENDIX

To evaluate improvements in signal-to-noise ratio (S/N), we employ the nadir-surface delay-time index identified within the clutter simulations available in the NASA Planetary Data System (Christoffersen *et al.* 2021). While using the first-return delay-time index is also possible, we found that the signal distribution of the first-return power extracted from the corresponding radargrams is often contaminated with noise and diffuse returns. Similarly, we find that the nadir-surface delay-time index predicted in the clutter simulations is offset by a few samples from the surface reflections in the radargrams. To correct this issue, we use the nadir-surface delay-time index as a center point and search around that index to find a local maximum power value within the radargrams using a buffer as described below (**Figures A1a** and **A1b**). Applying a buffered search tends to reduce noise contamination within our signal distribution and avoid "missing" the true maximum power (**Figures A1c**, **A1d**, and **A1f)**.

For the analyses presented in this article, we restrict the buffer size to 3 samples (0.1125 µs), resulting in a search gate of 7 samples (0.2625 µs). Once the optimized nadir-surface delay-time index is identified, we extract the corresponding power value from the radargram $(P_s)$. The noise $(P_n)$ is extracted 10 µs above the optimized nadir-surface delay-time index. For the histograms, the S/N is then calculated as:

$$S/N \;=\; 10\,log_{10}\!\left(\frac{P_s}{P_n}\right)$$

For the along-track S/N plots, we calculate the S/N using the average noise power value $(\bar{P}_n)$ across the entire radargram, which effectively stabilizes the result. However, since the S/N values vary greatly across the radargram (**Figure A1c**), we also apply a 201-sample (~93 km) smoothing window for the final plots (**Figures 4b**, **5b**, and **6b**).

The reported improvement in S/N of the VLR observations relative to 0° roll observations is determined by differencing the average along-track S/N values shown in **Figures 4b**, **5b**, and **6b**.

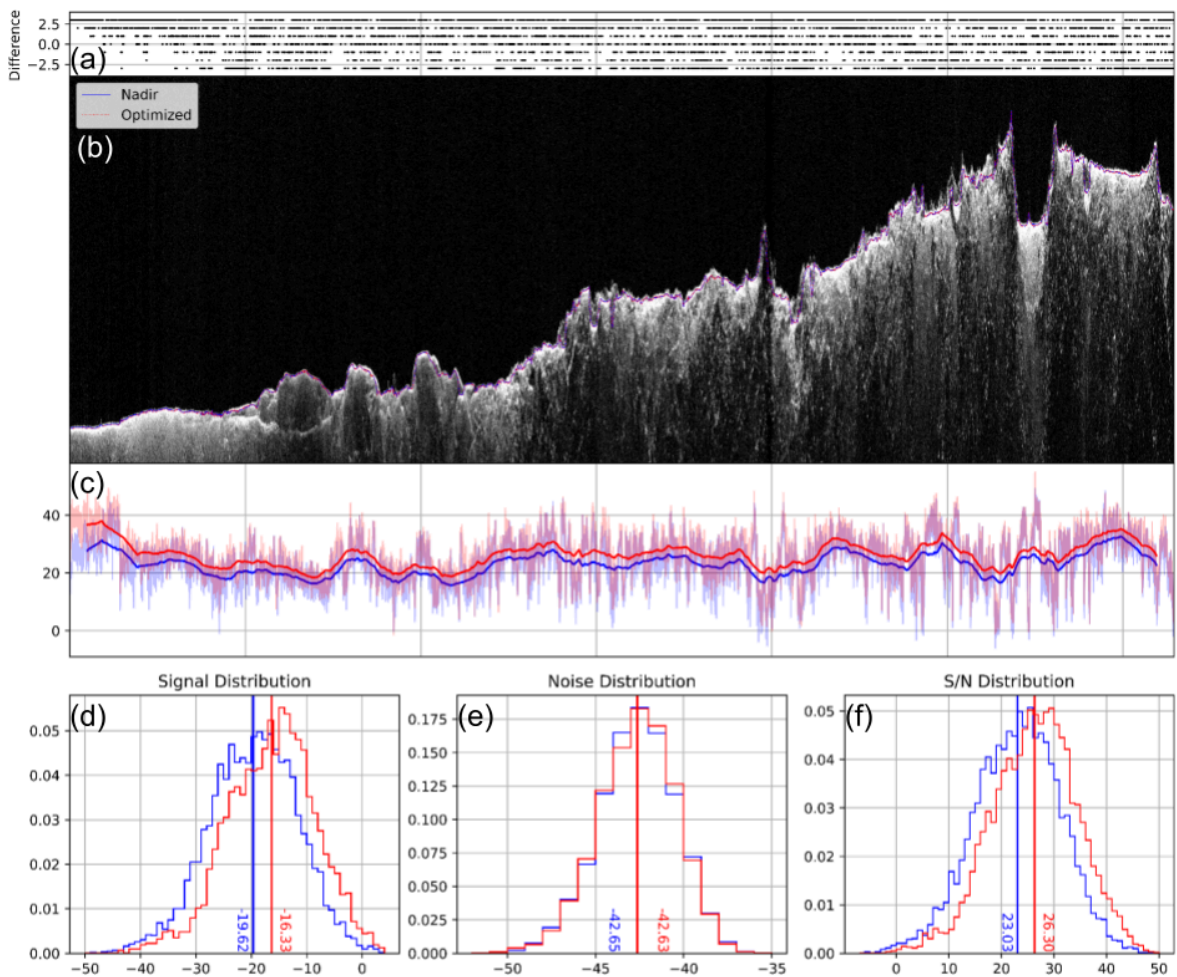

**Figure A1.** *Example radiometric workflow using 7858301 (VLR1). (a) Difference between the optimized nadir-surface delay-time index (using a 7-sample search gate) and the nadir-surface delay-time index available in the clutter simulation files. (b) Radargram of 7858301 focused on the surface return showing the original nadir delay time index (blue line) and the optimized index (red). (c) Along-track S/N plot with the raw values plotted in the background in muted colors and the smoothed results in the foreground with bold colors corresponding to the data already described. (d–f) Density histograms with means labeled for the (d) signal, (e) noise, and (f) S/N as extracted from the radargram with and without index optimization.*